\documentclass[aps,twocolumn,showpacs]{revtex4}
\usepackage{graphicx}   
\usepackage{latexsym}   

\newcommand{\beq}{\begin{equation}}
\newcommand{\eeq}{\end  {equation}}

\newcommand{\beqar}{\begin{eqnarray}}
\newcommand{\eeqar}{\end  {eqnarray}}

\newcommand{\bold}[1]{\mbox{\boldmath $#1$}}    
\newcommand{\smbold}[1]{\mbox{\boldmath\scriptsize $#1$}}
\newcommand{\zero}[1]{\hspace{0.1ex}\raisebox{1.0ex}{\scriptsize {$\circ$}}
  \hspace{-1ex}{#1}}
\newcommand{\del}{\partial}                     
\newcommand{\GeV}{{\rm GeV}}			
\newcommand{\MeV}{{\rm MeV}}                    
\newcommand{\fm}{{\rm fm}}                      
\newcommand{\eps}{\varepsilon}
\newcommand{\bfk}{\bold{k}}			
\newcommand{\bfr}{\bold{r}}			
\newcommand{\bfv}{\bold{v}}			
\newcommand{\rme}{{\rm e}}                      
\newcommand{\grad}{\bold{\nabla}}

\newcommand{\epsbar}{\bar{\eps}}
\newcommand{\pbar}{\bar{p}}
\newcommand{\rhobar}{\bar{\rho}}
\newcommand{\smk}{{\smbold{k}}} 		
\newcommand{\half}{\mbox{${1\over2}$}}          
\newcommand{\third}{\mbox{${1\over3}$}}         

\newcommand{\dthird}{\makebox{$d\over3$}}
\newcommand{\dhalf}{\makebox{$d\over2$}}

\begin{document}

\noindent{\sl Draft}

\hfill LBNL-00000

\title{Phase transition dynamics for baryon-dense matter}

\author{J{\o}rgen Randrup}

\affiliation{Nuclear Science Division, Lawrence Berkeley National Laboratory,
Berkeley, California 94720, USA}

\date{\today}

\begin{abstract}
We construct a simple two-phase equation of state 
intended to resemble that of compressed baryon-rich matter
and then introduce a gradient term in the compressional energy density
to take account of fintie-range effects in non-uniform configurations.
With this model we study the interface between the two coexisting phases
and obtain estimates for the associated interface tension.
Subsequently, we incorporate the finite-range equation of state
into ideal or viscous fluid dynamics 
and derive the collective dispersion relation 
for the mechanically unstable modes of bulk matter in the 
spinodal region of the thermodynamic phase diagram.
Combining these results with time scales extracted from existing 
dynamical transport simulations, we discuss the prospects for 
spinodal phase separation to occur in nuclear collisions.
We argue that these can be optimized by a careful tuning 
of the collision energy to maximize the time spent by the bulk of the system 
inside the mechanically unstable spinodal region of the phase diagram.
Our specific numerical estimates suggest cautious optimism 
that this phenomenon may in fact occur,
though a full dynamical simulation is needed for a detailed assessment.
\end{abstract}

\pacs{
25.75.-q,	
81.30.Dz,	
64.75.Gh,	
64.60.an 	
}

\maketitle

\section{Introduction}

The phase structure of strongly interacting matter presents a focal point 
for current theoretical and experimental investigations.
In particular, the Relativistic Heavy Ion Collider (RHIC) at BNL
is preparing for a beam energy scan 
that aims to identify signals of the expected critical point
and the CBM experiment at the future
Facility for Antiproton and Ion Research (FAIR) at GSI 
will explore the properties of compressed baryonic matter
and search for the expected first-order phase transition.

On the theoretical side, the situation is far from clear.
Whereas lattice QCD calculations \cite{AokiNature443,KarschPTP168} 
find that the deconfinement
phase transformation is of the crossover type at vanishing chemical potential,
$\mu=0$, they have inherent difficulties treating finite $\mu$ values and
any predictions in the baryon-rich domain are still very uncertain 
\cite{FodorPLB73,FodorJHEP04,GavaiPRD71}.
Even the very existence of a critical point has recently been called
into doubt \cite{ForcrandLAT2008}.
Experimental information would therefore be invaluable.

However, it will be no easy task to extract the thermodynamic phase structure 
from nuclear collision experiments.
In addition to the inherent problems arising from 
the smallness of the collision system 
(which renders its spatial configuration far from uniform)
and its rapid evolution 
(which prevents global equilibrium from being established),
the experimentalist is faced with the problem that there exists yet
no suitable dynamical model with which to simulate the collisions
for the purpose of anti\-cipating 
the observable effects of the phase structure.

This crucial point deserves elaboration:
First of all, the basic theory, quantum chromodynamics, is currently tractable
only in either the perturbative limit of hard elementary processes
or in the thermodynamic limit at vanishing (or small) net baryon density.
Any dynamical transport treatment of nuclear collisions
must therefore involve a considerable degree of modeling.

Ideally, one would devise a transport model that explicitly treats 
the dynamics of the microscopic degrees of freedom in the system, 
which change from being partonic in the deconfined sector 
to being hadronic in the confined sector.
Unfortunately, it has yet not been possible to develop such a description,
even for static scenarios.
Nevertheless, a variety of microscopic transport models
have achieved considerable success
with regard to calculating (and reproducing) observables for high-energy
collisions over a large range of energies.
However, their thermodynamic properties are (yet) inadequate.
For one thing, they usually lack detailed balance (as is often well justified
in the context of the dynamical processes for which they are intended) 
and therefore they are inadequate for thermal equilibrium.
Furthermore, these models do not (as of yet) incorporate 
a first-order phase transition.
Therefore, at this point, they appear to be unsuitable for simulations
that aim to bring out the dynamical effects of a phase transition and
elucidate their observability.

Considerable success has been obtained as well with
macroscopic models within the framework of fluid dynamics.
These models have the practical advantage that the underlying microscopic
degrees of freedom do not enter explicitly,
the state of the system being described merely through its local
energy and charge densities (and the associated currents)
with the interactions entering via the equation of state 
and the transport coefficients.
[Of course, in order to make contact with experiment,
such a treatment must ultimately convert the macroscopic information 
into hadrons by a suitable freeze-out prescription, 
but this occurs at densities well below the phase transition region
and is well developed.]
Thus fluid dynamics, especially ideal fluid dynamics for which the
transport coefficients vanish, posesses a very close relationship between 
the dynamics and the underlying object of study, the equation of state.

However, a closer analysis reveals that standard fluid dynamics has certain
inherent problems in the presence of a first-order phase transition.
Of particular importance is the fact that standard fluid dynamics 
is strictly local which leads to both static and dynamic shortcomings,
as we shall now discuss.

With regard to the former, imagine that two thermodynamically coexisting 
bulk systems are brought into contact along a common interface.
In a realistic description, a diffuse interface would develop,
with the various densities changing smoothly from one bulk value to the other,
and there would be an interface tension.
By contrast, when the equation of state is strictly local, 
the interface will be sharp (so the various densities will change abruptly 
from one bulk value to the other across the interface)
and there will be no interface tension.
Such a description would not be adequate for finite systems
such as blobs of matter produced in a collision,
whose sizes are determined primarily by the surface tension
and for which much of the matter is located in the diffuse surface region.
While this generic shortcoming may be less serious for high-energy collisions,
where the matter is being torn apart due to the rapid longitudinal stretching,
it is expected to play a significant role at the lower collision energies
relevant for the exploration of the deconfinement phase transition.

As for the dynamics, consider the evolution of nearly uniform matter
that has been prepared in a state of expansion 
at a density just above the phase coexistence region.
The system would then continue its expansion and the associated phase point
would soon enter the phase coexistence region in which uniform matter is
thermodynamically metastable.
While this would pose no particular problem as long as the deviations from
uniformity remain small, the further expansion would drive the phase point
into the region of spinodal instability, where uniform matter is both
thermodynamically and mechanically unstable (the speed of sound is imaginary).
As a result, density undulations would (and should) become amplified.

This scenario is familiar from many areas of physics
and it has been studied both theoretically
and experimentally for a variety of substances \cite{spinodal,PhysRep}.
Generally, the associated collective dispersion relation
(which in this situation gives the growth rate 
as a function of the wave number, $\gamma_k$) exhibits a maximum,
thus leading to preferential amplification of certain modes 
and the appearance of a characteristic length scale 
in the ensuing phase separation.
This remarkable phenomenon, known as spinodal phase decomposition,
is an indicator of a first-order phase transition.
It was found to present a powerful means for the experimental exploration 
of the nuclear liquid-gas phase transition \cite{BorderiePRL86,PhysRep},
because the unstable dilute bulk matter tends to condense into fragments
of similar sizes, a highly non-statistical outcome that is easy to identify
in the event analysis.
This success has given rise to the
hope that spinodal decomposition could be useful as well
for probing the confinement phase transition 
and some explorations of possible experimental signals have already been made
\cite{BowerPRC64,RandrupHIP22,KochPRC72,SasakiPRL99}.

Naturally, since standard fluid dynamics is local,
so is its collective dispersion relation, $\omega_k=v_0k$.
Consequently, inside the spindoal region of the phase diagram,
the growth rate will increase monotonically 
with the wave number of the undulation.
Thus $\gamma_k$ will not display a maximum and
the characteristic spinodal decomposition phenomenon would not be develop,
as density irregularities of ever smaller scale 
would be amplified at ever larger rates.
In ideal fluid dynamics, this problem would be computationally intractable 
(and in fact mathematically meaningless) \cite{RandrupPRL92}.
The inclusion of viscosity would modulate the dispersion relation
and cause the growth rate to approach a constant value for large $k$.
While this would facilitate the numerics,
the monotonic growth of $\gamma_k$ would still preclude
the occurrence of a spinodal decomposition.

Because of the considerable potential for fluid dynamics as a tool for
obtaining insight into the phase transition dynamics,
we address here this generic shortcoming.
Our main purpose is to illustrate the advantages of remedying this problem
and, along the way, make rough estimates for various key quantities.
Since the problem arises from the local nature of fluid dynamics,
we introduce a finite range into the treatment by means of a gradient term
in the compressional energy density.
The specific model developed here is intended to serve mainly as a framework
for illustrating the effect of incorporating a finite range into the
dynamcial desccription and the specific quantitative results
should be regarded as correspondingly rough.

We first (Sect.\ \ref{bulk}) construct a somewhat schematic equation of state
for uniform matter,
trying to incorporate the most essential features expected;
it should be considered as merely a temporary substitute subject to refinement.
Subsequently (Sect.\ \ref{grad}) the finite range is introduced
by means of a simple gradient term.
Then (Sect.\ \ref{surf}) we consider the equilibrium interface between 
two bulk systems, a property that could not be addressed with the standard
treatment due to its strict locality,
and obtain expressions for the associated interface tension.
The collective modes in bulk matter are then treated (Sect.\ \ref{coll})
and we consider particularly the spinodal growth rates
which now display the characteristic features known from other substances.
Finally (Sect.\ \ref{spin}), on this basis, 
we discuss the phase transition dynamics expected
for the planned nuclear collision experiments
and the prospects for spinodal decomposition to actually occur.

\section{ Bulk matter equation of state}
\label{bulk}

We wish to employ an equation of state that is suitable
for numerical illustrations.
For this purpose, we design a schematic model that is a generalization
of a classical gas in a density-dependent mean field.
The resulting equation of state has certain generic deficiencies
and the results should therefore not be taken at face value.
[For example, there are no bosonic degrees of freedom.]
It would of course be of interest to repeat the present analysis 
as more refined descriptions become available.

The equation of state provides the thermodynamic properties of bulk matter,
{\em i.e.}\ uniform matter of sufficient spatial extension to render 
finite-size effects (including those from any surfaces) insignificant.
In the microcanonical representation, the state of the system
is given in terms of the basic mechanical densities,
the (baryon) charge density $\rho$ and the energy density $\eps$
which we take as a thermal energy, $\kappa=\half d\rho T$,
plus a compressional energy, $w_0(\rho)$,
where $d$ is an adjustable parameter (equal to three for an ordinary gas).
Further adjustable parameters appear in the compressional energy density, 
$w_0(\rho)$, which is specified in App.\ \ref{mf}.

The key thermodynamic quantity is the entropy density $\sigma(\eps,\rho)$,
which we express in terms of the entropy density $\zero{\sigma}(\kappa,\rho)$
for a generalized ideal classical gas 
of density $\rho$ and thermal density $\kappa$,
\beq
\sigma(\eps,\rho)\ \equiv\ \zero{\sigma}(\eps-w_0(\rho),\rho)\ =\
\mbox{$5\over6$}d\rho-\third d\rho\ln{\rho\over\rho_T}\ ,
\eeq
where we have defiend the thermal density as
\beq
\rho_T(\eps,\rho)\equiv\left[{2\pi m\over h^2}T(\eps,\rho)\right]^{3/2}\ ,
\eeq
with $T(\eps,\rho)=\mbox{$2\over d$}[\eps-w_0(\rho)]/\rho$ (see below).
The Lagrange coefficients $\beta=1/T$ and $\alpha=-\mu/T$ are given by
\beqar
\beta(\eps,\rho)\! &\equiv&\! \del_\eps\sigma(\eps,\rho)
= \del_\kappa\zero{\sigma}(\kappa-w_0(\rho),\rho)
= \zero{\sigma}_\kappa,\\
\alpha(\eps,\rho)\! &\equiv& \!\del_\rho\sigma(\eps,\rho)
= \del_\rho\zero{\sigma}(\kappa-w_0,\rho)
= \zero{\sigma}_\rho-\zero{\sigma}_\kappa w_0',\,\,\
\eeqar
with $\zero{\sigma}_\kappa\equiv\del_\kappa\zero{\sigma}(\kappa,\rho)$,
$\zero{\sigma}_\rho\equiv\del_\rho\zero{\sigma}(\kappa,\rho)$,
and $w_0'\equiv\del_\rho w_0(\rho)$.
Thus, the temperature and the chemical potential are
\beqar
T(\eps,\rho) &=& {1\over\beta}\ =\ {2\over d}\,{\eps-w_0(\rho)\over\rho}\
=\  {2\over d}\,{\kappa\over\rho}\ ,\\
\mu(\eps,\rho) &=& -\alpha T\ 	
	=\ \third dT\ln{\rho\over\rho_T}+w_0'(\rho)\ .
\eeqar
The pressure and the enthalpy density may be obtained subsequently,
\beqar
p(\eps,\rho) &=& \sigma T-\eps+\mu\rho\ =\
\third d\rho T -w_0 +\rho w_0'\ ,\\
h(\eps,\rho) &\equiv& p+\eps\ =\ \mbox{$5\over6$}d\rho T+\rho w_0'\ .
\eeqar

Two bulk systems with the densities $(\eps_1,\rho_1)$ and $(\eps_2,\rho_2)$
are in mutual thermodynamic equilibrium iff the total entropy
is stationary under arbitrary exchanges of energy and charge,
yielding the requirement that they have equal temperatures,
chemical potentials, and pressures:
$\beta_1\doteq\beta_2\equiv\beta_0$, $\alpha_1\doteq\alpha_2\equiv\alpha_0$, 
$p_1\doteq p_2\equiv p_0$.
Thus phase coexistence requires that the gradient of $\sigma(\eps,\rho)$,
$(\sigma_\eps,\sigma_\rho)\equiv(\del_\eps\sigma,\del_\rho\sigma)$,
be the same at the two phase points and, furthermore
(since $p=T[\sigma-\beta\eps-\alpha\rho]$), 
that the tangent to $\sigma(\eps,\rho)$ at these two points be common.
Furthermore, local thermodynamic stability requires that 
the second variation of the entropy be positive under such exchanges,
yielding the requirement that the curvature matrix of $\sigma(\eps,\rho)$
be positive definite.
Consequently, the region of spinodal instability is delineated by 
the occurrence of a vanishing curvature eigenvalue.

In the canonical representation $\eps$ is replaced by $T$,
and the free energy density is then of special interest,
\beqar\nonumber
f_T(\rho) &\equiv& \eps_T(\rho)-T\sigma_T(\rho)\ 
	=\ \mu_T(\rho)\rho-p_T(\rho)\\
	&=& \rho T\ln{\rho\over\rho_T}-\rho T+w_0(\rho)\ ,
\eeqar
where the subscript $T$ indicates that the quantity is obtained at
the specified temperature.
We also note that the slope of the free energy density
is the chemical potential, $\del_\rho f_T(\rho)=\mu_T(\rho)$.

\begin{figure}          
\includegraphics[angle=0,width=3.1in]{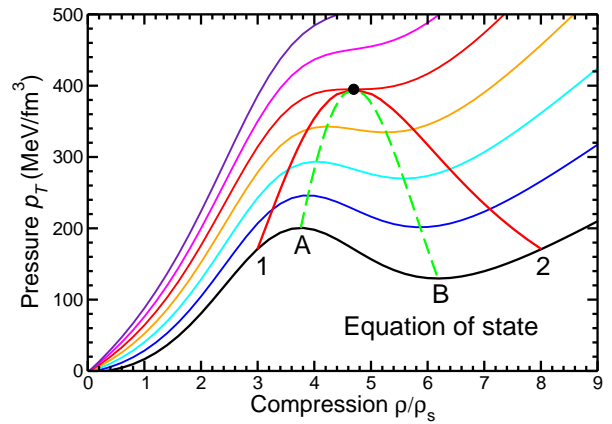}	
\caption{The equation of state $p_T(\rho)$:
The pressure $p$ as a function of the density $\rho$ 
for a range of temperatures, $T/T_c=0,\mbox{$1\over4$},\mbox{$1\over2$},
\mbox{$3\over4$},1,\mbox{$5\over4$},\mbox{$3\over2$}$,
obtained with the adopted model.
The phase coexistence (solid)
and the spinodal (dashes) boundaries are indicated;
they coincide at the critical point (dot).
}\label{f:EoS}
\end{figure}            

\begin{figure}          
\includegraphics[angle=0,width=3.1in]{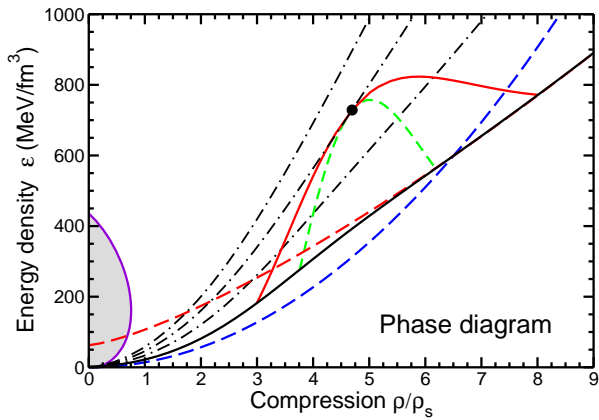}	
\caption{The phase diagram in the $\rho$\,-\,$\eps$ plane,
as obtained from the equation of state (Fig.\ \ref{f:EoS}),
with the phase coexistence boundary (solid, red),
the (isothermal) spinodal boundary (short dashes, green),
and the critical point (dot) indicated.
The hadronic freezeout line  (lower left)
is included for reference (from Ref.\  \cite{RandrupPRC74}).
Also shown are the two functions $w_H(\rho)$ (dashed blue curve)
and $w_Q(\rho)$ (dashed red curve) between which
the compressional energy $w_0(\rho)$ (solid curve) is interpolated,
as well as three isentropic phase trajectories (dot-dashed curves),
for which $\rho\delta\eps\! =\! (\eps\!+\!p)\delta\rho$.
}\label{f:rho-eps}
\end{figure}            

\begin{figure}          
\includegraphics[angle=0,width=3.1in]{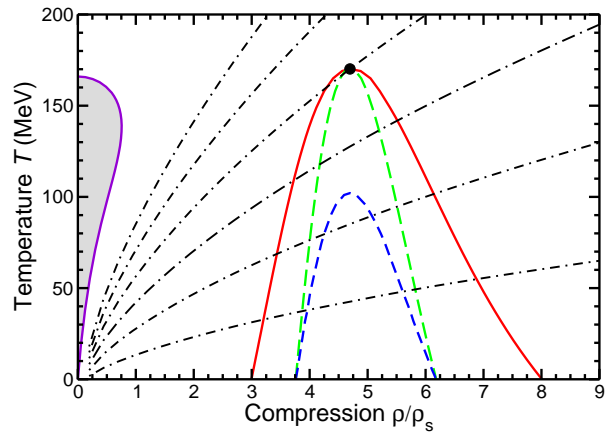}	
\caption{The phase diagram in the $\rho$\,-\,$T$ plane indicating
the phase coexistence boundary (solid),
the isothermal spinodal boundary (long dashes),
the isentropic spinodal boundary (short dashes),
and the critical point (dot).
Also shown are several isentropic phase trajectories,
for which $\rho\delta\eps\! =\! (\eps\!+\!p)\delta\rho$.
}\label{f:rho-T}
\end{figure}

For $w_0(\rho)$ we use an interpolated form
that produces a first-order phase transition (see App.\ \ref{mf}).
To achieve a semi-quantative correspondence with expectations for
actual baryon-rich matter,
we adjust the parameters such that the coexistence densities at $T=0$
are $\rho_1=3\rho_0$ and $\rho_2=8\rho_0$,
where $\rho_0\approx0.153\,\fm^{-3}$ is the nuclear saturation density;
the associated zero-temperature specific heat is then
$w_0(\rho_2)-w_0(\rho_1)=590\,\MeV/\fm^3$.
Furthermore, the value $d=5.5$ yields
a critical temperature of $T_c=170\,\MeV$;
the critical density is then $\rho_c=4.70\,\rho_0$.
Other valus of particular interest are listed in Table \ref{t:1}.
(These values are of course somewhat arbitrary
but will serve well for illustrative purposes.)
The resulting equation of state, $p_T(\rho)$, is shown in Fig.\ \ref{f:EoS},
while Fig.\ \ref{f:rho-eps} displays the associated phase diagram
expressed in terms of the mechanical densities $\rho$ and $\eps$.
The more familiar $(\rho,T)$ phase diagram,
for which the energy density $\eps$ has been replaced by the temperature $T$,
is shown in Fig.\ \ref{f:rho-T}.
It is important to recognize that whereas the transformation from $\eps$ to $T$
is always unique, the reverse transformation is triple-valued 
in the presence of a phase transition:
Any $(\rho,T)$ phase point inside the phase coexistence region 
in Fig.\ \ref{f:rho-T} could arise 
any of three different $(\rho,\eps)$ phase points in Fig.\ \ref{f:rho-eps}.

At a given temperature $T$, bulk matter at the two different densities 
$\rho_1$ and $\rho_2$ are in mutual thermodynamic equilibrium
if the corresponding tangents of $f_T(\rho)$ are common:
the two chemical potentials are then equal 
since $\mu_T(\rho)=\del_\rho f_T(\rho)$,
and the relation $p_T(\rho)=\mu_T(\rho)\rho-f_T(\rho)$ guarantees
that also the two pressures match.
Thus phase coexistence at $T=0$ requires that the tangents of $w_0(\rho)$ 
at the two densities be common.
[We have used this propertiy to guide our choice of mean field.]
As the temperature is increased, the difference between 
the two coexistence densities will steadily shrink
until they coincide at the critical temperature $T_c$.

At supercritical temperatures ($T\!>\!T_c$)
the pressure increases steadily with $T$, $\del_\rho p_T>0$,
whereas its behavior is undulatory at subcritical temperatures:
when the density is increased from the lower to the higher coexistence density,
the pressure exhibits first a maximum and then a minimum.
The associated densities $\rho_A$ and $\rho_B$ where $p_T(\rho)$ is stationary
delineate the region of mechanical instability, 
within which $\del_\rho p_T(\rho)$ is negative.
Since $\del_\rho p_T(\rho)=\dthird T+\rho w_0''$,
the spinodal boundary densities at $T=0$ are determined by 
$\rho w_0''(\rho)\doteq0$ and $w_0''(\rho)$ is negative in between.
The region of mechanical instability shrinks steadily as $T$ is increased
and disappears at $T_c$, which is thus determined by the condition
$\dthird T_c+\rho_m w_0''(\rho_m)\doteq0$, where $\rho_m$ is the density 
at which $\rho w_0''(\rho)$ is most negative.
The adopted compressional energy density $w_0(\rho)$ 
is depicted in Fig.\ \ref{f:EoS}
together with the coexistence and spinodal boundaries.

Dynamical transport calculations suggest that the expansion stage
in a nucleus-nucleus collision proceeds in an approximately isentropic manner
\cite{ArsenePRC75}, 
{\em i.e.}\ the entropy per (net) baryon remains nearly constant.
Since $T\rho^2\delta(\sigma/\rho)=
(\rho\delta\eps-\mu\rho\delta\rho)-(h\delta\rho-\mu\rho\delta\rho)
=\rho\delta\eps-h\delta\rho$,
the isentropic trajectories in the $(\rho,\eps)$ phase plane
are characterized by $\rho\delta\eps\doteq h\delta\rho$.
Figs.\ \ref{f:rho-eps} and \ref{f:rho-T} display several such 
isentropic phase trajectories and they are seen to not be noticeably affected 
by the presence of the phase transition.
This feature brings out the fact that the locations of the boundaries for
thermodynamic and mechanical instability, including the critical point itself, 
result from a rather subtle interplay between the underlying interactions.
One may therefore expect that the overall phase evolution obtained
in a dynamical transport calculation is not very sensitive to the specific
phase structure.

\section{Gradient corrections}
\label{grad}

The above thermodynamics discussion applies to bulk matter, 
{\em i.e.}\ large and uniform systems.
In heavy-ion physics, the systems encountered are neither
and it is therefore practically important to extend the treatment
to systems whose densities vary with the location,
$\tilde{\eps}(\bfr)$ and $\tilde{\rho}(\bfr)$,
where we use a tilde over a quantity as a reminder that
it pertains to a non-uniform system.

As a simple way to take approximate account of finite range effects,
we employ a gradient correction in the compressional energy.
(A gradient term was also employed in recent hydrodynamical studies of
the hadron-quark first-order phase transition \cite{Skokov}.)
Accordingly, we write the local interaction-energy density on the form
\beqar
\tilde{w}(\bfr) &=& w_0(\tilde{\rho}(\bfr))\
	+\ \half C(\grad\tilde{\rho}(\bfr))^2 \\ \nonumber
&=&	w_0(\tilde{\rho}(\bfr))\
	+\ \half a^2\eps_{\rm g}
	\left({\grad\tilde{\rho}(\bfr)\over\rho_{\rm g}}\right)^2\ .
\eeqar
It is convenient to write the strength of the gradient term
on the form $C=a^2\eps_{\rm g}/\rho_{\rm g}^2$, 
where $\rho_{\rm g}$ is a characteristic charge density
and $\eps_{\rm g}$ is a characteristic energy density.
Since we are here particularly interested in the dynamics
in the phase transition region, we choose the phase point 
$(\rho_{\rm g},\eps_{\rm g})$ to be in the middle of the
phase coexistence region,
$\rho_{\rm g}\doteq\rho_c=4.70\,\rho_0$ and
$\eps_{\rm g}\doteq\eps_{T=T_c/2}(\rho_c)=
\half(w_0(\rho_c)+\eps_c)=561\,\MeV/\fm^3$.
The strength of the gradient term is then governed by the length $a$
which we consider to be somewhat adjustable.
Our present calculations have been made with $a=0.2\,\fm$.

The introduction of the gradient term leads to gradient corrections
in the expressons for the various thermodynamic quantities.
In order to derive those, we start from the entropy density,
which we assume to still have the form
$\tilde{\sigma}(\bfr)=\zero{\sigma}(\tilde{\kappa}(\bfr),\tilde{\rho}(\bfr))
	=\zero{\sigma}(\tilde{\eps}(\bfr)-\tilde{w}(\bfr),\tilde{\rho}(\bfr))$,
where $\tilde{\kappa}(\bfr)=\tilde{\eps}(\bfr)-\tilde{w}(\bfr)$ 
is the local thermal energy density.
A variation of the total entropy 
$S[\tilde{\eps}(\bfr),\tilde{\rho}(\bfr)]=\int\!d\bfr\,\tilde{\sigma}(\bfr)$
then yields the local Lagrange coefficients 
$\tilde{\beta}$ and $\tilde{\alpha}$,
\beqar\label{beta}
\tilde{\beta}(\bfr) &\doteq& {\delta S\over\delta\tilde{\eps}(\bfr)}\
	=\ \zero{\sigma}_\kappa(\tilde{\kappa}(\bfr),\tilde{\rho}(\bfr))\ 
	=\ 1/\tilde{T}(\bfr)\ ,\\ \label{alpha} \nonumber
\tilde{\alpha}(\bfr) &\doteq& {\delta S\over\delta\tilde{\rho}(\bfr)}\
	=\ \zero{\sigma}_\rho(\tilde{\kappa}(\bfr),\tilde{\rho}(\bfr))
	-\tilde{\beta}(\bfr)w_0'(\tilde{\rho}(\bfr))\\
	&+&C\grad(\tilde{\beta}(\bfr)\cdot\grad\tilde{\rho}(\bfr))\
	=\ -\tilde{\mu}(\bfr)/\tilde{T}(\bfr)\ .
\eeqar
Using that the entropy density gradient is then given by
\beq
\grad\tilde{\sigma}(\bfr) = \tilde{\beta}(\bfr)\grad\tilde{\eps}(\bfr)
	+\tilde{\alpha}(\bfr)\grad\tilde{\rho}(\bfr))
	-C\grad(\tilde{\beta}(\bfr)(\grad\tilde{\rho}(\bfr))^2) ,
\eeq
we see that the following expression for the local pressure,
\beq\label{p}
\tilde{p}(\bfr)\ =\ \tilde{\sigma}(\bfr)\tilde{T}(\bfr)
	-\tilde{\eps}(\bfr)+\tilde{\mu}(\bfr)\tilde{\rho}(\bfr)
	+C(\grad\tilde{\rho}(\bfr))^2\ ,
\eeq
leads to the relation
\beq
\grad{\tilde{p}(\bfr)\over\tilde{T}(\bfr)}\
=\	-\tilde{\eps}(\bfr)\grad\tilde{\beta}(\bfr)
	-\tilde{\rho}(\bfr)\grad\tilde{\alpha}(\bfr)\ ,
\eeq
which can be regarded as a generalization of the familiar
thermodynamic relation $\delta(p/T)=-\eps\delta\beta-\rho\delta\alpha$.
This relation ensures that $\tilde{p}(\bfr)$ will be constant 
whenever $\tilde{T}(\bfr)$ and $\tilde{\mu}(\bfr)$ are.
We also note that the gradient correction to the compressional energy
migrates directly into the free energy density,
\beqar\nonumber
\tilde{f}_T(\bfr) &=& \kappa_T(\tilde{\rho}(\bfr))+\tilde{w}(\bfr)
-T\zero{\sigma}(\kappa_T(\tilde{\rho}(\bfr)),\tilde{\rho}(\bfr))\ ,\\
&=& f_T(\tilde{\rho}(\bfr))+\half C(\grad\tilde{\rho}(\bfr))^2\ .
\eeqar

\section{Interface equilibrium}
\label{surf}

Once the finite-range effects have been included in the thermodynamics,
one may treat the interface between two coexisting phases.
For this purpose, we consider a semi-infinite geometry 
with the two coexisting systems
having a planar interface perpendicular to the $x$ direction.
The coexistence values of temperature, chemical potential, and pressure
are denoted by $T_0$, $\mu_0$, and $p_0$.

We first note that global equilibrium, 
including equilibrium between two bulk systems with a common interface,
requires that the total entropy $S$ be constant under variations 
$\delta\tilde{\eps}(x)$ and $\delta\tilde{\rho}(x)$
that conserve the total energy $E=\int dx\,\tilde{\eps}(x)$ 
and the total (net) charge $B=\int dx\,\tilde{\rho}(x)$,
\beqar\nonumber
0 &\doteq& \delta S-\beta_0\delta E-\alpha_0\delta B\\
&=& \delta\!\int\!dx\,[\tilde{\sigma}(x)
	-\beta_0\tilde{\eps}(x)-\alpha_0\tilde{\rho}(x)]\\ \nonumber
&=&	\int\!dx\,\left\{[\tilde{\beta}(x)-\beta_0]\delta\tilde{\eps}(x)
		+[\tilde{\alpha}(x)-\alpha_0]\delta\tilde{\rho}(x)\right\}\ ,
\eeqar
thus implying spatial constancy of the temperature and the chemical potential,
$\tilde{\beta}(x)\doteq\beta_0$ and $\tilde{\alpha}(x)\doteq\alpha_0$,
as one should expect.

Therefore, assuming that the temperature is constant,
$\tilde{\beta}(x)=\beta_0=1/T_0$,
it is convenient to work in the canonical framework and
our analysis is then similar to that carried out by Ravenhall {\em et al.}\
\cite{RavenhallNPA407}.
With the temperature given,
the local density $\tilde{\rho}(x)$ determines the local energy density, 
$\tilde{\eps}(x)=\dhalf\tilde{\rho}(x)T_0+\tilde{w}(x)$, 
and the local entropy density is then in turn determined, $\tilde{\sigma}(x)
=\zero{\sigma}(\tilde{\eps}(x)-\tilde{w}(x),\tilde{\rho}(x))
=\zero{\sigma}(\dhalf\tilde{\rho}(x)T_0,\tilde{\rho}(x))$.
The local free energy density is then readily obtained,
\beq 
\tilde{f}(x) = \tilde{\eps}(x)-T_0\tilde{\sigma}(x)
	= f_{T_0}(\tilde{\rho}(x))+\half C(\del_x\tilde{\rho}(x))^2\ ,
\eeq
where $f_T(\rho)$ is the free energy density in bulk matter
at temperature $T$ and density $\rho$ (see Sect.\ \ref{bulk}).
The corresponding bulk chemical potential is $\mu_T(\rho)=\del_\rho f_T(\rho)$,
while the bulk pressure is $p_T(\rho)=\mu_T(\rho)\rho-f_T(\rho)$.

\begin{figure}          
\includegraphics[angle=0,width=3.1in]{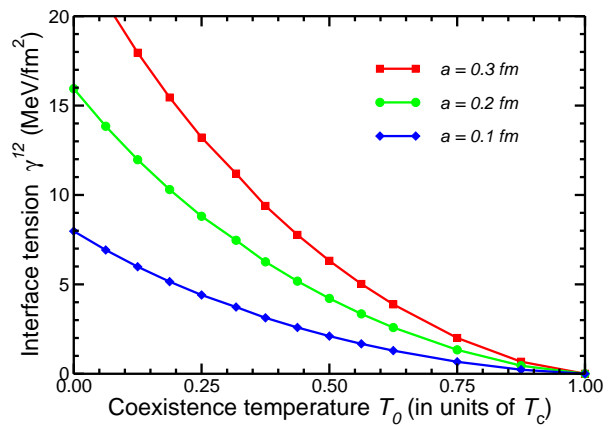}	
\caption{The specific interface tension $\gamma_{T_0}^{12}$
as a function of the coexistence temperature $T_0$
for various values of the range $a$.
}\label{f:gamma12}
\end{figure}            

The condition for equilibrium can now be expressed as
\beqar
0 &\doteq& \delta\!\int\! dx\left[\tilde{f}(x)-\mu_0\tilde{\rho}(x)\right]\\ 
\nonumber &=&	\int\!dx
	\left[\mu_{T_0}(\tilde{\rho}(x))-C\del_x^2\tilde{\rho}(x)
	-\mu_0\right]\delta\tilde{\rho}(x)\ ,
\eeqar
which then requires
\beq\label{eq}
C\del_x^2\tilde{\rho}(x)\ \doteq\ \mu_{T_0}(\tilde{\rho}(x))-\mu_0\
	=\ \del_\rho\Delta f(\tilde{\rho}(x))\ .
\eeq
Here $\Delta f(\rho)$ is the difference between the free energy density
of a uniform system of density $\rho$, $f_{T_0}(\rho)$, 
and the corresponding  ``Maxwell'' free energy density,
defined as the free energy density along the common tangent,
\beq
f_{T_0}^M(\rho)\ \equiv\ f_{T_0}(\rho_i)+\mu_0(\rho-\rho_i)\ 
\leq\ f_{T_0}(\rho)\ ,
\eeq
where $\rho_i$ refers to either one of the two coexistence densities.
Thus $\Delta f(\rho)$ can be thought of as the free energy (density) 
gained by performing a phase mixture.

\begin{figure}
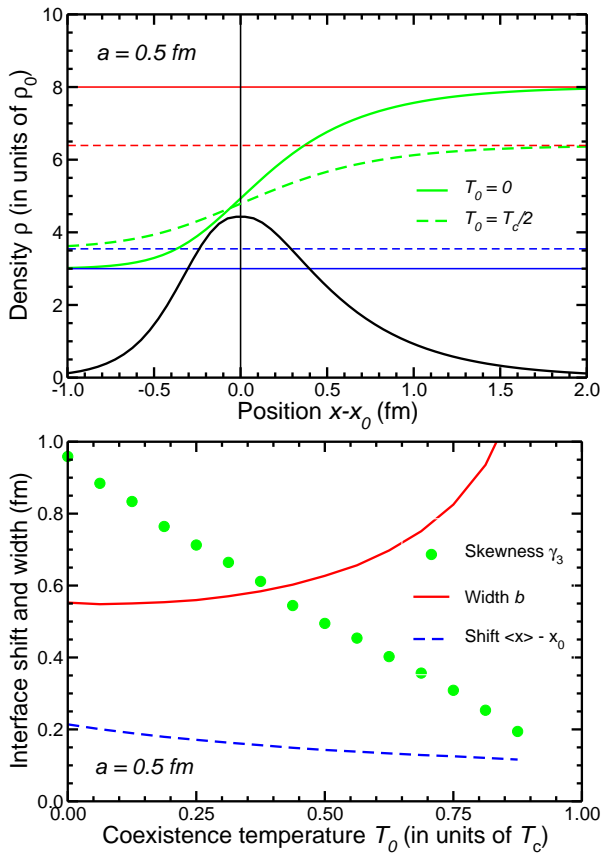
          
\includegraphics[angle=0,width=3.1in]{Fig5a}
\includegraphics[angle=0,width=3.1in]{Fig5b}
\caption{{\em Top:} The surface profile $\tilde{\rho}(x)$ 
for $T=0$ (solid) and $T=\half T_c$ (dashed), using as a reference for $x$
the position $x_0$ where the chemical potential equals the coexistence value
of the bulk chemical potential, $\mu_T(\tilde{\rho}(x=x_0))\doteq\mu_0(T)$.
The limiting (coexistence) densities are shown by the horizonthal lines,
while the bottom curve is the interface location function $g(x)$ for $T=0$.
{\em Bottom:} The mean location $\bar x$ of the interface and its width $b$
as functions of temperature.
Also shown is the profile skewness parameter
$\gamma_3\equiv\langle(x-\bar{x})^3\rangle/b^3$.
}\label{f:rho}
\end{figure}            

The equilibrium condition (\ref{eq}) for the density profile $\tilde{\rho}(x)$
is formally equivalent to an equation of motion for a particle of mass $C$
moving in the potential $V(\rho)=-\Delta f(\rho)$, with $\rho$ denoting the
coordinate and $x$ the time.
(We note that $\Delta f(\rho)$ vanishes at the two coexistence densities
and is positive in between.)
Conservation of the corresponding energy $\half C(\del_x\tilde{\rho})^2+V$
(which vanishes) then determines the gradient at each position,
\beq\label{rhox}
\del_x\tilde{\rho}(x)\ 
	=\ \left[{2\over C}\Delta f(\tilde{\rho}(x))\right]^{1\over2}\ .
\eeq

The local excess in the free energy density due to the interface
(see App.\ \ref{interface}) is given by
\beqar
\tilde{f}^{12}_{T_0}(x) &=& \tilde{f}(x)-f_{T_0}^M(\tilde{\rho}(x))\\ \nonumber
	&=& \Delta f(\tilde{\rho}(x))+\half C(\del_x\tilde{\rho}(x))^2\
	=\ 2\Delta f(\tilde{\rho}(x))\ .
\eeqar
The total deficit in free energy per unit interface area,
equal to the interface tension, is then given by 
\footnote{There are two common notations for the interface tension,
$\sigma$ and $\gamma$; 
since $\sigma$ might be confused with the entropy density, we use $\gamma$,
hoping that it will not be confused with the spinodal growth rate.}
\beqar
\gamma^{12}_{T_0} &=& \int_{-\infty}^{+\infty}dx\,\tilde{f}^{12}_{T_0}(x)\
=\ 2\!\int{d\tilde{\rho}(x)\over\del_x\tilde{\rho}(x)}\,
\Delta f(\tilde{\rho}(x))\\
\nonumber
&=& \int_{\rho_1}^{\rho_2} d\rho\left[2C\Delta f(\rho)\right]^{1\over2}\
=\ a\int_{\rho_1}^{\rho_2} {d\rho\over\rho_{\rm g}}
\left[2\eps_{\rm g}\Delta f(\rho)\right]^{1\over2}\ .
\eeqar
We note that this quantity can be obtained without explicit knowledge 
of the interface density profile $\tilde{\rho}(x)$
and it scales directly with the length parameter $a$.
It is shown in Fig.\ \ref{f:gamma12} as a function of temperature.
As expected, it decreases steadily from its maximum value at $T=0$
until it vanishes at $T_c$.
With the (somewhat arbitrary) parameter values adopted,
the zero-temperature interface tension is $\gamma_0^{12}\approx16~\MeV/\fm^3$,
about 16 times the familiar nuclear surface tension.
This value lies near the lower end of the rather wide range of expected 
values for the tension between quark and nuclear matter
(typical low values are $10\!-\!20\,\MeV/\fm^2$,
while typical high values are $50\!-\!100\,\MeV/\fm^2$,
see for example Refs.\ \cite{HeiselbergPRL70,VoskresenskyNPA723}).

The density profile itself, $\tilde{\rho}(x)$,
can be obtained by integrating Eq.\ (\ref{rhox}),
\beq
\tilde{\rho}(x)\ =\ \tilde{\rho}(x_0) +\rho_c\int_{x_0}^x
\left[{2\over\eps_g}\Delta f(\tilde{\rho}(x))\right]^{1\over2}{dx\over a}\ ,
\eeq
where $x_0$ is some location where the density is known.
We note that it would not be feasible to start the integration
at $x_0\to\pm\infty$, where $\tilde{\rho}(x_0)\to\rho_i$,
since the gradient vanishes in the same limit, $\tilde{\rho}_x(x_0)\to0$.
In stead, we take $x_0$ to be that location
where the function $\Delta f(\tilde{\rho}(x))$ has its maximum.
Since the derivative $\del_\rho\Delta f(\rho)=\mu_{T_0}(\rho)-\mu_0$
thus vanishes at $\rho=\tilde{\rho}(x_0)$, 
it follows that the local bulk chemical potential at $x_0$ 
matches the coexistence value, $\mu_{T_0}(\tilde{\rho}(x_0))=\mu_0$, and this
relation can be used to find the starting density value $\tilde{\rho}(x_0)$.
[The local bulk chemical potential $\mu_{T_0}(\rho)$ must match
the coexistence value $\mu_0=\mu_{T_0}(\rho_i)$ for some intermediate density 
because $\mu_{T_0}(\rho)$ exhibits an undulation between $\rho_1$ and $\rho_2$,
going first through a maximum $\mu_{T_0}(\rho_A)>\mu_0$
and then through a minimum $\mu_{T_0}(\rho_B)<\mu_0$,
so it must equal $\mu_0$ somewhere between $\rho_A$ and $\rho_B$.]

The density profile $\tilde{\rho}(x)$ is shown in Fig.\ \ref{f:rho}
for $T_0=0$ and $T_0=\half T_c$.
At each temperature, it scales horizontally with the length parameter $a$.
The interface profile can be characterized by the cumulants
of the associated {\em interface location function},
$g(x)=\tilde{\rho}_x(x)/(\rho_2-\rho_1)$ (see App.\ \ref{interface}).
So the mean interface location is 
$\bar{x}=\langle x\rangle\equiv\int dx xg(x)$,
while its width $b$ is the corresponding dispersion,
$b^2=\langle(x-\bar{x})^2\rangle$.
A convenient measure of the profile skewness is given by the dimensionless
parameter $\gamma_3\equiv\langle(x-\bar{x})^3\rangle/b^3$.
As the temperature is increased,
the profile grows progressively wider and more symmetric,
while its mean location moves closer to $x_0$.
With the adopted parameter values
we find $\bar{x}-x_0=1.01\,a$, $b=2.62\,a$, and $\gamma_3=0.96$ at $T_0=0$.

It should be noted that in the present simple treatment,
where the finite range is taken into account by means of a gradient term,
the interface tension as well as the detailed density profile shape
reflect the specific density dependence of the free energy $f_T(\rho)$,
{\em i.e.}\ they follow directly from the employed bulk equation of state,
apart from scalings related to the strength of the gradient term.

\section{Collective modes}
\label{coll}

We now wish to study the dynamical response
to the introduction of small density undulations 
imposed on a system that is static and uniform,
$\delta\eps(\bfr)=\tilde{\eps}(\bfr)-\epsbar$ and 
$\delta\rho(\bfr)=\tilde{\rho}(\bfr)-\rhobar$.
We first note that the local change in the pressure is then of a similar form,
$\delta p(\bfr)=\tilde{p}(\bfr)-\bar{p}$
with $\bar{p}=p(\epsbar,\rhobar)$.
For simplicity, we assume that the time evolution 
is described by fluid dynamics
and we first disregard dissipation.
The equations of motion then arise from
energy-momentum conservation, $\del_\mu T^{\mu\nu}=0$,
together with conservation of (baryon) charge, $\del_\mu j^\mu=0$.

Assuming that the local flow velocities $\bfv(\bfr)$ are non-relativistic,
we may ignore $v^2$ and thus put $\gamma$ to unity.
This yields the following five equations of motion,
\beqar
0&=&\del_\mu T^{\mu0}(\bfr,t)\ 
	\approx\ \del_t\delta\eps +\bar{h}\del_iv^i\ ,\\
0&=&\del_\mu T^{\mu i}(\bfr,t)\
	\approx\ \bar{h}\del_tv^i +\del^i\delta p\ ,\\
0&=&\del_\mu j^\mu(\bfr,t)\,\,\,\
	\approx\ \del_t\delta\rho +\bar{\rho}\del_iv^i\ ,
\eeqar
where $\bar{h}=\epsbar+\pbar$ is the enthalpy density of the uniform system.
As usual, the equations for $T^{\mu\nu}$
can be combined to a sound-wave equation, 
while a comparison of the first and last equations yields the evolution 
of the density disturbance in terms of that of the energy disturbance, so
\beqar\label{eom}
\del_t^2\delta\eps(\bfr) &=& \del_i\del^i\delta p(\bfr)\ ,\\ \label{rhoeps}
\bar{h}\,\del_t\delta\rho(\bfr) &=& \bar{\rho}\,\del_t\delta\eps(\bfr)\ .
\eeqar

It is straightforward to see that, 
to leading order in the disturbances $\delta\eps(\bfr)$ and $\delta\rho(\bfr)$,
the local pressure is
\beq
\tilde{p}(\bfr)\ \approx\ p(\tilde{\eps}(\bfr),\tilde{\rho}(\bfr))
-C\rhobar\nabla^2\rho(\bfr)\ .
\eeq
The first term is the usual local-density approximation,
{\em i.e.}\ the pressure is calculated as in uniform matter 
that has been prepared with the local density values,
while the second term arises from the gradient correction 
to the chemical potential (\ref{alpha}).
Therefore, to the same order,
\beq
\nabla^2\delta p(\bfr)\ \approx\
	p_\eps\nabla^2\eps(\bfr)+p_\rho\nabla^2\rho(\bfr)
	-C\rhobar\nabla^4\rho(\bfr)\ ,
\eeq
where $p_\eps\equiv\del_\eps p(\eps,\rho)$ 
 and  $p_\rho\equiv\del_\rho p(\eps,\rho)$ 
evaluated at the local phase point
$(\eps,\rho)=(\tilde{\eps}(\bfr),\tilde{\rho}(\bfr))$.

If we require the undulations to be of harmonic form,
$\delta\eps(\bfr)=\eps_\smk\exp(i\bfk\cdot\bfr-i\omega t)$ and
$\delta\rho(\bfr)=\rho_\smk\exp(i\bfk\cdot\bfr-i\omega t)$,
then Eq.\ (\ref{rhoeps}) requires $\bar{h}\rho_\smk=\rhobar\eps_\smk$.
The dispersion relation is then readily obtained from (\ref{eom}),
\beq
\omega_k^2\ =\ v_s^2k^2+C{\rhobar^2\over\bar{h}}k^4\
=\ v_s^2k^2+a^2{\eps_{\rm g}\over\bar{h}}{\rhobar^2\over\rho_{\rm g}^2}k^4\ .
\eeq
Here the first term is what emerges in ordinary ideal fluid dynamics, 
with $v_s$ being the isentropic speed of sound
(see Eq.\ (\ref{vs})),
\beq\label{v02}
v_s^2\ =\ p_\eps +{\rhobar\over\bar{h}}p_\rho\
= -{\bar{T}\over\bar{h}}\left[\bar{h}^2\sigma_{\eps\eps}
	+2\bar{h}\rhobar\sigma_{\eps\rho}+\rhobar^2\sigma_{\rho\rho}\right]\ ,
\eeq
with 
$\sigma_{\eps\rho}\equiv\del_\eps\del_\rho\sigma(\eps,\rho)$
evaluated at $(\eps,\rho)=(\epsbar,\rhobar)$, {\em etc.}.
This part of the dispersion relation is perfectly linear, $\omega_k=v_sk$.
That pathological behavior is modified by the gradient term
which generally increases $\omega_k^2$.
In the spinodal region, where $v_s^2$ is negative, 
the collective frequency is imaginary, $\omega_k=\pm i\gamma_k$,
and the gradient term then suppresses the growth of high-$k$ modes.
As a result, the growth rate $\gamma_k$ will exhibit a maximum
followed by a rapid fall-off to zero as a function of the wave number $k$,
as is familiar from other substances exhibiting spinodal instability
\cite{spinodal, PhysRep}.

It is instructive to write the growth rate on the form
$\gamma_k=|v_s|k(1-k^2/k_{\rm max}^2)^{1/2}$
where the maximum wave number for which spinodal instability occurs 
is given by
\beq
k_{\rm max}^2\ =\ {\bar{h}\over C}{|v_s^2|\over\rhobar^2}\ =\
-{\bar{h}\over\eps_{\rm g}}{\rho_{\rm g}^2\over\rhobar^2}{|v_s^2|\over a^2}\ .
\eeq
The maximum in $\gamma_k$ occurs at the ``optimal'' wave number 
$k_{\rm opt}=k_{\rm max}/\sqrt{2}$ and,
as the amplification process proceeds,
undulations of this size will become dominant
and a characteristic spinodal pattern will thus emerge.
The corresponding largest growth rate is
$\gamma_{\rm opt}=\half|v_s|k_{\rm max}=|v_s|k_{\rm opt}/\sqrt{2}$.
This quantity scales inversely with the length parameter $a$ %
and the associated optimal wave length
$\lambda_{\rm opt}=2\pi/k_{\rm opt}$ thus scales directly with $a$. 
Consequently, 
an increase of $a$ will increase the scale of the most
rapidly amplified mode as well as the associated shortest growth time 
$t_{\rm opt}=1/\gamma_{\rm opt}$.

\begin{figure}
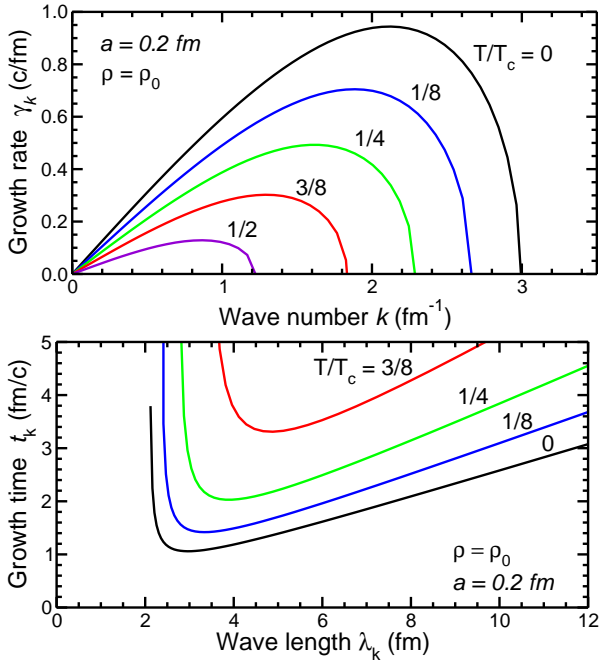
          
\includegraphics[angle=0,width=3.1in]{Fig6a}	
\includegraphics[angle=0,width=3.1in]{Fig6b}	
\caption{The growth rate $\gamma_k$ vs.\ the wave number $k$ ({\em top}) and
the corresponding growth times $t_k=\hbar/\gamma_k$
vs.\ the wave length $\lambda_k=2\pi/k$ ({\em bottom}), 
for various temperatures $T$ at $\rho=\rho_0$.
}\label{f:tk}
\end{figure}            

\begin{figure}
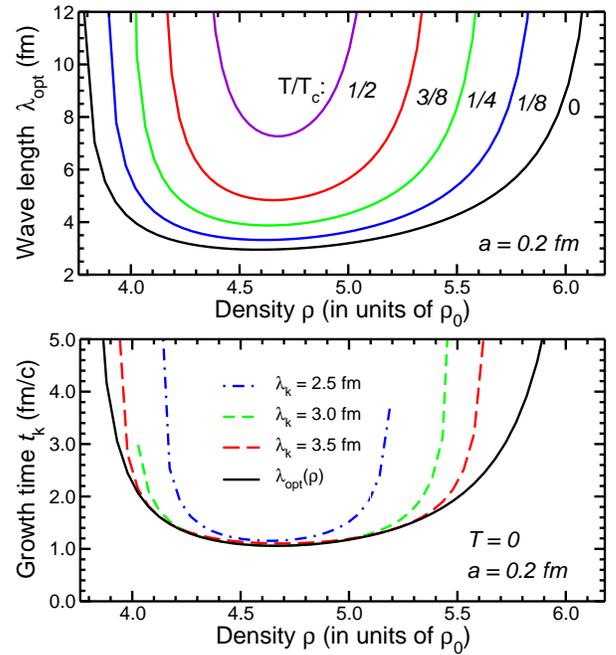
          
\includegraphics[angle=0,width=3.1in]{Fig7a}	
\includegraphics[angle=0,width=3.1in]{Fig7b}	
\caption{The optimal wave length $\lambda_{\rm opt}$ ({\em top})
for various temperatures $T$ 
and the growth time $t_k=\hbar/\gamma_k$ ({\em bottom}) at $T=0$
for several values of the wave length $\lambda_k=2\pi/k$,
as functions of the degree of compression, $\rho/\rho_0$.
}\label{f:trho}
\end{figure}

The spinodal growth rates $\gamma_k$ depend on the environment,
as specified for example by $\rhobar$ and $\bar{T}$.
The temperature dependence is illustrated in Fig.\ \ref{f:tk} 
for $\rhobar=\rho_c$, while the density dependence 
is shown in Fig.\ \ref{f:trho}.
$\gamma_k(\rhobar,\bar{T})$ generally vanishes along the spinodal boundary
and it decreases as a function of temperature.
With the present model, we thus find that the fastest mode at $\rhobar=\rho_c$
has a wave length of $\lambda_{\rm opt}\approx3\,\fm$
and a growth time of $t_{\rm opt}\approx1.0\,\fm/c$.
As the temperature is raised, 
the maximum wave number $k_{\rm max}$ decreases
as do the optimal values $k_{\rm opt}$ and $\gamma_{\rm opt}$.
While the obtained temperature dependence is quite significant,
it should be recognized that the thermal properties 
of the present model may not be realistic.
(By contrast, the spinodal growth rates in nuclear matter
are relatively independent of temperature at the low end
because of the fermion nature of the constituents \cite{PhysRep}.)
On the other hand,
the dependence of $\gamma_k(\rhobar,\bar{T})$ on density is more moderate
in the phase region of most rapid growth (as in dilute nuclear matter
\cite{PhysRep}).

It is important to appreciate that 
the phase region of instability for ideal fluid dynamics
is bounded by the {\em isentropic} spinodal (where $v_s=0$)
and it therefore lies inside the region of thermodynamic instabilty
which is bounded by the {\em isothermal} spinodal
(see Fig.\ \ref{f:rho-T}).
There are unstable isentropic modes whenever $v_s^2<0$.
Insertion of the susceptibilities 
$\sigma_{\eps\eps}$, $\sigma_{\eps\eps}$, $\sigma_{\eps\eps}$
(see App.\ \ref{kappa}) into Eq.\ (\ref{v02}) 
yields an explicit expression for the speed of sound,
\beq
v_s^2\ \equiv\
{\rho\over h}\left({\del p\over\del\rho}\right)_{s\equiv\sigma/\rho}\
=\ {\mbox{$5\over9$}dT+\rho w_0'' \over\mbox{$5\over6$}dT+w_0'}\ ,
\eeq
so the condition for instability becomes $\mbox{$5\over9$}dT+\rho w_0''<0$.
At zero temperature this amounts to $\del_\rho^2 w_0(\rho)<0$,
which occurs exactly within the isothermal spinodal density region,
as one would expect since $T=0 \Leftrightarrow\sigma=0$.
However, as $T$ is increased, the region of isentropic instability shrinks
faster than the region of isothermal instability and it disappears entirely
at  $T_{\rm max}=\mbox{$3\over5$}T_c$.

The above analysis was based on ideal fluid dynamics which conserves entropy,
$\del_\mu\sigma^\mu=0$, 
where $\sigma^\mu=\sigma u^\mu$ is the entropy current density.
We wish to conclude this section by briefly discussing
the effects of including viscosity into the fluid-dynamic treatment.
Within the non-relativistic framework used for the derivation of the
dispersion relations for the normal modes in bulk matter,
the inclusion of shear and bulk viscosity into the fluid-dynamic treatment
changes the pressure gradient by the term
$-\grad[\mbox{$4\over3$}\eta+\zeta]\grad\cdot\bfv$
where $\eta$ and $\zeta$ are the shear and bulk viscosity coefficients,
respectively. The dispersion equation is then modified accordingly,
\beq
\omega^2\ =\ v_s^2k^2	
	+a^2{\eps_{\rm g}\over\bar{h}}{\rhobar^2\over\rho_{\rm g}^2}k^4
	-i[\mbox{$4\over3$}\eta+\zeta]{k^2\over\bar{h}}\omega\ ,
\eeq
where we have assumed that the combination 
$\xi\equiv\mbox{$4\over3$}\eta+\zeta$ can be regarded as constant, 
for simplicity.
Clearly, the zero-frequency modes occur for the same wave numbers as before,
$k=0$ and $k=k_{\rm max}$,
and the inclusion of viscosity
does not afffect the location of the spinodal boundary.
(However, if thermal conductivity were included,
the spinodal boundary would gradually expand towards the isothermal boundary
\cite{RavenhallNPA407}.)
To leading order, the viscosity adds a negative imaginary term 
to the frequency, $-\mbox{$i\over2$}\xi k^2/\bar{h}$,
which in turn gives rise to an exponential damping factor.
Furthermore, inside the spinodal region the collective frequencies
are still purely imaginary, $\omega=i\gamma_\pm$, and we find
\beq
\gamma_\pm\ =\ \pm\left[|v_s^2|k^2
	-a^2{\eps_{\rm g}\over\epsbar}{\rhobar^2\over\rho_{\rm g}^2}k^4
	+\mbox{$1\over4$}\xi^2{k^4\over\bar{h}^2}\right]^{1\over2}
	-\half\xi{k^2\over\bar{h}}\ .
\eeq
Thus the growth rate $\gamma_+$ is reduced by $\approx\half\xi k^2/\bar{h}$
and the optimal wave number becomes smaller as well.
Even though the qualitative features will remain the same,
the viscous effects may be quantitatively important 
\cite{Skokov}.

In order to illustrate the key role played by the finite range
in producing spinodal decomposition,
let us briefly consider what would happen without the gradient term.
We already noted that the resulting non-viscous dispersion relation
would exhibit linear growth, $\gamma_k=|v_s^2|k$, and thus not
favor any particular length scale.
When viscosity is included, the growth rate would still grow
monotonically, $\del_k\gamma_+(k)>0$, but level off for large $k$,
\beq
\gamma_+(k\to\infty)\ \approx\ {\rhobar\over\xi}|v_s^2|\left[
1-\mbox{$3\over4$}{\rhobar^2\over\xi^2}{|v_s^2|\over k^2}+\dots\right]\ .
\eeq
Thus the large-$k$ divergence characteristic of standard ideal fluid
dynamics would be eliminated, but there would still not be a preferred
length scale.

\section{Spinodal decomposition?}
\label{spin}

Using the particular model and parameter values chosen here,
we now discuss the prospects for spinodal decomposition to occur
during a nuclear collision.
Although the explored mean-field model is rather simplistic
and the specific parameter values are somewhat uncertain,
the resulting features appear to be within the range of plausibility.
It may therefore be instructive to explore the consequences.
Obviously, as further progress is made, both theoretically and experimentally,
the models should be appropriately refined.

In order to understand under what experimental conditions
spinodal decomposition may actually occur,
it is useful to consider how the thermodynamic conditions
in the bulk of the collision system evolve in the course of time.
Such phase trajectories were studied for gold-gold collisions
with a variety of existing dynamical transport models \cite{ArsenePRC75}
and we shall make use of those results for our estimates.
Ref.\ \cite{ArsenePRC75} calculated the evolution of 
the mechanical phase point $(\rho(t),\eps(t))$ in order to avoid
making any assumption about local thermalization;
since we are here mainly concerned with the expansion stage,
we assume that equilibrium has been established and so
we shall frame our discussion in terms of the canonical phase variables 
$(\rho(t),T(t))$ which are somewhat more intuitive.

\begin{figure}          
\includegraphics[angle=0,width=3.1in]{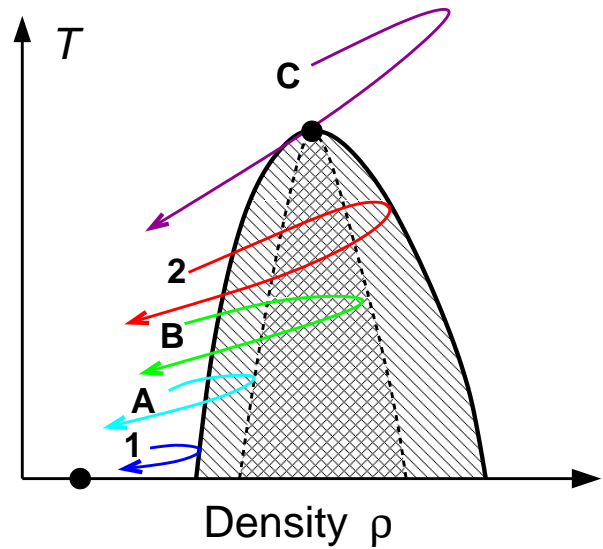}	
\caption{Illustration of the dynamical phase traejctories for the
most compressed matter produced at the various threshold collision energies 
$E_1$, $E_A$, $E_B$, $E_2$, $E_c$.
}\label{f:E}
\end{figure}

Generally speaking, 
the prospects for spinodal decomposition can be expected to be better
the more time the bulk of the matter spends inside 
the region of spinodal instability.
Let us therefore consider how this quantity develops with the collision energy.
For the discussion below, we assume that the equation of state
has the expected form
with a first-order phase transition terminated by a critical point,
as drawn schematically in Fig.\ \ref{f:E} (see also Fig.\ \ref{f:rho-T}).
It seems natural to introduce a number of threshold values
of the collision energy $E$: $E_1$, $E_A$, $E_B$, $E_2$, $E_c$.
Their meaning is illustrated in Fig.\ \ref{f:E} 
and they will be explained in turn below.

At the lowest collision energies, $E<E_1$,
the compressions achieved are insufficient to bring any part of the matter 
inside the region of phase coexistence.
Consequently, at such low energies, 
it would probably not be possible to probe the phase transition.

As the collision energy is raised above $E_1$, 
the phase trajectory $(\rho(t),T(t))$ of the most compressed matter
makes ever larger incursions into the phase coexistence region.
Characterizing such phase trajectories by the highest compression achieved,
$\rho_{\rm max}(E)$, we expect this ``turning point'' 
to gradually move across the phase coexistence region as $E$ is raised.
It first enters the spinodal region for $E=E_A$ and it has traversed it fully
for $E=E_B$, reaching the other side of the coexistence region at $E=E_2$.

At collision energies above $E_2$, the steady expansion of the bulk matter
subsequent to its maximum compression drags its phase trajectory through 
the phase coexistence region (and the spinodal region within it).
As $E$ is increased the slope of the expansion phase trajectory steepens
(see Ref.\ \cite{ArsenePRC75}) and the traversal time becomes steadily shorter,
both because the expansion is faster and because the region of instability
becomes narrower at the ever higher excitations encountered.
At a certain ``critical''collision energy, $E=E_c$,
the phase trajectory passes right through the critical point $(\rho_c,T_c)$
and at supercritical collision energies, $E>E_c$,
the phase trajectory will miss the unstable phase region altogether.

Generally, the evolving local thermodynamic conditions during a collision
will differ from one location to another.
Consequently, a single collision event gives rise to an entire bundle
of phase trajectories and
the above discussion pertains to just the phase trajectory
of the most compressed matter of the collision system which,
for a symmetric collision, is presumably located around the center.
Furthermore, there is a dependence on the geometric features
of the collision system, such as the nuclear sizes and the impact parameter.
Thus the precise meaning of the various threshold energies is somewhat fuzzy 
and they play primarily a conceptual role.
This underscores the fact that quantitative predictions
must rely on detailed dynamical calculations.

Our special interest here concerns the relatively narrow 
interval of collision energy within which the turning point 
lies inside the spinodal region, $E_A<E<E_B$.
Intuitively, one would expect that collision energies slightly below $E_B$
would be optimal for maximizing the time spent by the phase trajectory
inside the spindoal region.
Such collisions, in turn, would presumably be most favorable for the
development of spinodal decompostion.
The transport calculations reported in Ref.\ \cite{ArsenePRC75}
suggest that this optimal beam energy is $5-15~\GeV$ per nucleon
for a stationary target setup.
The presice values depend not only on the specific location of the
spinodal phase boundaries, {\em i.e.}\ on the specific equation of state
(which is still unknown),
but also on the complications arising from the non-uiformity of the density
and its time evolution.

In order to get a rough idea of the degree of spinodal growth
that could be expected during a given collision,
we assume that we know the time evolution of the bulk density, $\rho(t)$,
and the associated temperature, $T(t)$.
As pointed out above, these quantities are local and the present analysis
employs suitable average values that can be taken as representative of an
extended part of the system.
For a given collision energy between $E_A$ and $E_B$,
the phase trajectory $(\rho(t),T(t))$ enters the spinodal region 
at the time $t=t_i$ and exits it again at the time $t=t_f$.
Thus, for $t_i<t<t_f$ the collective dispersion relation
yields unstable modes with associated growth rates 
$\gamma_k(t)\equiv\gamma_k(\rho(t),T(t))$ 
which serve to amplify irregularities in the density.

An accurate calculation would need to take account not only of the
distribution of density fluctuations but also of the fact the entire
scenario changes in time (and relatively rapidly).
This is beyond our present scope and we seek to obtain a simple estimate
by considering the following amplification coefficient \cite{RandrupPRL92},
\beq
\Gamma_0\ \equiv\ \int_{t_i}^{t_f}\gamma_0(t)\,dt\
	\approx\ \mbox{$2\over3$} \gamma_{\rm max} \Delta t\ ,
\eeq
where $\gamma_0(t)$ is the maximum growth rate at the time $t$,
$\gamma_0(t)\equiv\gamma_{\rm opt}(\rho(t),T(t))$
and $\gamma_{\rm max}$ is the largest growth rate overall.
The factor of two thirds accounts roughly for the fact that $\gamma_0(t)$
has a parabola-like appearance,
starting out from zero at $t_i$, 
exhibiting a broad maximum of $\gamma_{\rm max}$,
and then dropping to zero again at $t_f$,
so we put $\langle\gamma_0(t)\rangle\approx\mbox{$2\over3$}\gamma_{\rm max}$.
Judging from the transport calculations reported in Ref.\ \cite{ArsenePRC75},
we estimate the duration of the spinodal stage to be 
$\Delta t\equiv t_f-t_i\approx6\,\fm/c$.
To estimate $\gamma_{\rm max}$ is more difficult.
Our present calculations give an overall fastest growth time of 
$\approx1\,\fm/c$, obtained for relatively broad range of densties
and for zero temperature.
However, the compression achieved in a nuclear collision
is inevitably accompanied by a corresponding agitation,
and we therefore expect $T/T_c=\third-\half$ to be more realistic.
In this connection it should be realized that the the rapid decrease 
of the calculated growth rate as the temperature is increased
is to some extent a reflection of the fact that the present instability region
is bounded the isentropic rather than the isothermal spinodal line.
On the other hand, the inclusion of dissipation
(which would expand the region of instability)
is expected to slow the dynamics down to a significant degree \cite{Skokov}.
Therefore it is probably more realistic to expect the fastest growth time
to be several times that most optimistic value,
so we use $\gamma_{\rm max}^{-1}\approx2-4\,\fm/c$

With these rough numbers, we then find the value of the
amplification coefficient to be
$\Gamma_0\approx\mbox{$2\over3$}\!\cdot\!6/(2-4)=1-2$.
The corresponding amplitude growth factor \cite{RandrupPRL92}
is then given by $G_0\equiv\exp(\Gamma_0)\approx2.7-7.4$.
When trying to judge the significance of this value,
one should keep in mind that the density-density correlation function
is proportional to the {\em square} of the amplitude growth factor,
{\em i.e.}\ $\langle\delta\rho(\bfr_1)\exp(i\bfr_{12}\!\cdot\!\bfk)
	\delta\rho(\bfr_2)\rangle\sim G_k^2$.

It should also be realized that after the system exits the spinodal
instability region of the phase diagram, it has still to traverse
the metastable region between the spinodal boundary and the phase
coexistence line.  
While nearly uniform matter is mechanically stable in this regime,
this is no longer so for matter having significant deviations from uniformity.
Therefore, the undulations resulting from even relatively modest amplifications
during the spinodal stage may be further amplified during the metastable
stage and thus lead to observationally interesting clumping of 
the expanding matter.

The above numerical estimates were obtained for the adopted range value
of $a=0.2\,\fm$ which is of course rather uncertain.
We recall that it leads to an interface tension of 
$\gamma^{12}_{T=0}\approx16\,\MeV/\fm^2$ and an optimal wavelength of 
$\lambda_{\rm opt}\approx4\,\fm$ at $T\approx\third T_c$.
The interface tension is at the lower end of what has been used by various 
authors \cite{HeiselbergPRL70,VoskresenskyNPA723}, 
but does not appear to be unreasonable
considering the large uncertainties on this quantity.
As for the wave length, it is of interest to note that
a spherical volume having such a diameter would contain a baryon number 
of $B_0=\mbox{$\pi\over6$}\lambda_0^3\rho_c\approx24$.
At a temperature of half the critical value,
the completion of the phase decomposition would distribute
this matter approximately evenly between the two coexisting phases,
assuming the high-density phase is concentrated in a sphere
embedded into the low-density phase.
Such a blob of deconfined matter is large enough
to constitute a macroscopic statistical source,
while at the same time being probably a sufficiently small part 
of the total system to permit the simultaneous formation of several such blobs
and thus make it feasible to perform a size-correlation analysis.

If we were to use only half that range, $a=0.1\,\fm$, 
those quantities would decrease correspondingly to 
$\gamma^{12}_{T=0}\approx8\,\MeV/\fm^2$,
which would be somewhat low in comparison to the existing estimates,
though perhaps not impossible,
and $\lambda_{\rm opt}(\third T_c)\approx2\,\fm$,
leading to $B_0\approx3$ which seems too small 
to constitute a macroscopic source that could have observational significance.
On the other hand, if we were to double the range, $a=0.4\,\fm/c$,
we would obtain $\gamma^{12}_{T=0}\approx32\,\MeV/\fm^2$,
which would seem quite reasonable,
but $\lambda_{\rm opt}(\third T_c)\approx8\,\fm$ (hence $B_0\approx192$)
would be far too large to produce a useful effect.

This analysis reveals that whether a given model leads to
spinodal phase decomposition in simulations of nuclear collisions 
depends rather delicately on the specific values of its parameters.
Consequently, it is far from certain that collisions of real nuclei 
would in fact produce this phenomenon at any collision energy.

The estimates above suggest that collisions within a suitably tuned energy
range may produce bulk matter that stays inside the mechanically unstable 
phase region sufficiently long for some degree of spinodal clumping to occur.
However, it is hard to predict whether the amplification 
of the fluctuations will suffice to bring about the characteristic 
spinodal enhancement of a certain length scale.
Indeed, the degree of amplification might be so marginal
that the matter will revert to approximate uniformity after reentering 
the stable regime and no clumping-like phase separation would then occur.

This uncertainty underscores the need for studying 
the observable consequences of a spinodal decomposition.
Indeed, only if the spinodal phenomenon manifests itself in detectable signals
can it be turned into a useful tool for probing the equation of state.
It is as of yet far from clear whether any proposed ``signals'' of the phase 
decomposition would in fact survive the subsequent expansion stage
dominated by hadronic resonances.
While current miscroscopic transport models present useful tools for such
investigations, the insight that can be gained will likely remain somewhat 
limited until we achieve a better understanding of the
phase transition dynamics itself.

\section{Concluding remarks}

The present study was motivated by the need for developing theoretical models
that can address the dynamics of the confinement phase transition that is
expected to occur during the expansion stage 
of collisions between heavy nuclei at suitably tuned energies.
As a step towards this goal, we have developed a simple model within which 
we have studied both the collective dispersion relation for the
mechanically unstable modes of bulk matter in the 
spinodal region of the thermodynamic phase diagram
{\em and} the properties of the interface between the two coexisting phases
into which such an ustable system seeks to decompose.
These properties are central to the phase transition dynamics
and since they are fundamentally related 
it is important they that be treated consistently.
As far as we are aware, this is the first time that these different properties
have been addressed within the same model framework.

The key element is the inclusion of a finite interaction range
without which there would be neither an interface tension
nor spinodal decomposition.
Indeed, a zero-range model would render any interface perfectly sharp
and there would be no associated energy cost,
hence no basis for determining the geometric structure
of a system composed of coexisting phases 
(a zero-range model would admit even fractal intermingling of the phases).
Furthermore, without a finite range to suppress the dynamics of short 
wavelength disturbances, 
the growth rate would increase steadily with wave number
(even in the presence of viscosity), 
hence not dispay a maximum as is characteristic of spinodal decomposition.
Thus, any model that aspires to be of use for phase transition dynamics
must incorporate a finite range.

A recent study by Skokov and Voskresensky has sought to accomplish that
by means of a gradient term in the free energy functional \cite{Skokov}.
In the present study, we take a similar approach by
constructing an equation of state in which the interaction-energy density
in non-uniform matter contains a gradient term.
This can be thought of as approximating a convolution with a kernel
of finite range, 
as is often done for the mean field in low-energy nuclear physics.

Considering that no quantitatively reliable calculations are yet available,
we have had to fix the parameters of the equation of state for bulk matter
on the basis of our best guess for the location of the coexistence region,
including the critical point.
Consequently, our results should not be considered as more than suggestive.
The additional range parameter entering into the gradient term
has been adjusted to yield reasonable values for the interface tension
and the spinodal growth rates.
Importantly, a change in this range by a factor of two or more
would render the model results either implausible 
(relative to existing estimates)
or phenomenologically uninteresting 
(in that the resulting model would not produce spinodal decomposition
in a collision scenario).

With the gradient term included, we have then studied the equilibrium
interface between two coexisting phases and determined the temperature
dependence of the density profile and the associated interface tension.

In order to address the collective modes in bulk matter,
a dynamcial model is needed and we have adopted fluid dynamics,
which can readily be adapted to the finite-range equation of state.
The gradient term suppresses the growth of short wavelengths and 
thus yields a physically reasonable dispersion relation for the spinodal modes.

Taking guidance from existing phase trajectories
extracted from various transport simulations \cite{ArsenePRC75}, 
we have used the calculated growth rates to estimate the degree of
amplification that might occur when the collision energy is adjusted
to maximize the exposure to the spinodal instabilities.
The resulting amplification amounts to one or two factors of $e$,
which may suffice to trigger a phase separation
due to the subsequent further amplification from the 
intermediate metastable phase region.
While this conclusion gives grounds for guarded optimism, 
it also brings out the fact that a full dynamical simulation is needed 
for a more detailed assessment.


\section*{Acknowledgements}
We wish to acknowledge many helpful discussions with P.F.~Bedaque, B.~Friman,
U.~Heinz, V.~Koch, R.~Sharma, W.J.~Swiatecki, and D.N.\ Voskresensky.
This work was supported by the Director, Office of Energy Research,
Office of High Energy and Nuclear Physics,
Nuclear Physics Division of the U.S. Department of Energy
under Contract No.\ DE-AC02-05CH11231.

\appendix
\section{Compressional energy}
\label{mf}

The key quantity in the employed illustrative equation of state
is the energy density associated with the compression
of bulk matter at zero temperature, 
$w_0(\rho)=\eps_{T=0}(\rho)$.
We obtain this basic function by interpolating 
between a ``hadron gas'' and a ``quark-gluon plasma'',
\beq
w_0(\rho)\ =\ \chi(\rho)w_H(\rho)\ +\ [1-\chi(\rho)]w_Q(\rho)\ ,
\eeq
where the interpolation function is taken as
\beq
\chi(\rho) = [1+\rme^{(\rho-\rho_\chi)/\rho_w}]^{-1}\ 
\eeq
and each phase is taken to display a simple power form,
\beq
w_H(\rho) = c_H(\rho/\rho_0)^2,\,\,\
w_Q(\rho)\ =\ c_Q(\rho/\rho_0)^{4/3}+B ,
\eeq
$\rho_0\approx0.153\,\fm^{-3}$ being the nuclear saturation density.

For a wide range of parameter values,
the resulting compressional energy density $w_0(\rho)$ 
exhibits a region of negative curvature,
thus ensuring the existence of a first-order phase transition.
[We recall that phase coexistence requires equal chemical potentials,
$\mu(\rho_1)\doteq\mu(\rho_2)$,
hence equal slopes of $w_0(\rho)$ (since $\mu_{T=0}=\del_\rho w_0$),
as well as equal pressures, $p(\rho_1)=p(\rho_2)$,
hence a common tangent of $w_0(\rho)$
(since $p_{T=0}=\mu\rho-f=\rho\del_\rho w_0-w_0$).]
Adopting the specific values 
$c_H=92.6$, $c_Q=288.9$, and $B=408.3$ (all in $\MeV/\fm^3$)
together with $\rho_\chi=3.2535\,\rho_0$ and $\rho_w=0.945\,\rho_0$,
we obtain $\rho_1=3\rho_0$ and $\rho_2=8\rho_0$
for the zero-temperature coexistence densities.
(It would be straightforward to obtain other values
by readjusting the parameters.)

The resulting compressional energy density $w_0(\rho)$ 
is shown in Fig.\ \ref{f:EoS} by the solid (black) curve,
while the dashed (blue and red) curves are the individual functions
$w_H(\rho)$ and $w_Q(\rho)$.
It is apparent from the plot that the coexistence features
depend delicately on the parameter values.
One would therefore expect any model calculation of the phase structure
to be endowed with rather large uncertainties,
thus reinforcing the need for experimental information.

Table \ref{t:1} summarizes the values of the various quantities of interest
at the phase and spinodal boundaries:

\begin{table}[h]
\begin{tabular}{c|ccccc}
\hline
~\#~&~$T\,(\MeV)$& $\rho/\rho_0$ & $\eps\,(\MeV\!/\fm^3)$ 
	& ~$p\,(\MeV\!/\fm^3)$ & ~$\mu\,(\MeV)$ \\
\hline
1	& 0	& 3.00	& 182	& 171	& 769	\\
A	& 0	& 3.76	& 276	& 201	& 829	\\
c	& 170	& 4.70	& 729	& 395	& 923	\\
B	& 0	& 6.18	& 563	& 130	& 732	\\
2	& 0	& 8.01	& 772	& 171	& 769	\\
\hline
\end{tabular}
\caption{Values of the temperature $T$, compression $\rho/\rho_0$,
energy density $\eps$, pressure $p$, and chemical potential $\mu$
at the two zero-temperature coexistence points ($\#1$ and $\#2$),
the two spinodal boundaries at zero temperature ($\#A$ and $\#B$),
and the critical point ($\#c$) (see Figs.\ \ref{f:EoS} and \ref{f:rho-eps}).
}\label{t:1}\end{table}

.
\section{Susceptibilities}
\label{kappa}

Mechanical stability is determined by the curvature tensor  $\bold{\sigma}$
of the entropy density, which in the present model has the elements
$\sigma_{\eps\eps}$, $\sigma_{\rho\eps}=\sigma_{\eps\rho}$, 
$\sigma_{\rho\rho}$, where
\beqar\label{see}
\sigma_{\eps\eps}\! &\equiv&\! \del_\eps^2\sigma(\eps,\rho)
= -\dhalf{\rho\over(\eps-w_0)^2}\ =\ -\dhalf{\rho\over\kappa^2}\ <\ 0 ,\\
\sigma_{\rho\eps}\! &\equiv&\! \del_\rho\del_\eps\sigma(\eps,\rho)\
= \sigma_{\eps\rho} 
= \dhalf{1\over\kappa} +\dhalf{\rho w_0'\over\kappa^2}\ ,\\ \label{ser}
\sigma_{\rho\rho}\! &\equiv&\! \del_\rho^2\sigma(\eps,\rho)
= -\mbox{$5\over6$}{d\over\rho} 
-d{w_0'\over\kappa}
-\dhalf{\rho w_0''\over\kappa}
-\dhalf{\rho w_0'^2\over\kappa^2} .\,\,\,\,\,\,\ \label{srr}
\eeqar
with $\kappa(\eps,\rho)\equiv\eps-w_0(\rho)=\half d\rho T$.
We generally have 
\beq\label{Dsigma}
|\bold{\sigma}|\ \equiv\ 
\sigma_{\eps\eps}\sigma_{\rho\rho}-\sigma_{\eps\rho}^2\
=\ -{\sigma_{\eps\eps}\over\rho T}\,\del_\rho p_T(\rho)\ ,
\eeq
which in the present case amounts to
\beq
|\bold{\sigma}|\ =\ 
	{2\over d}\,{\third dT+\rho w_0''\over\rho^2T^3}\ .
\eeq
Since the occurrence of mechanical instability requires that
at least one of the eigenvalues of $\bold{\sigma}$ be positive,
and $|\bold{\sigma}|$ is the product of the eigenvalues,
it follows that mechanical instability occurs at densities for which
$w_0''(\rho)$ is negative and then extends up to the temperature
$T_{\rm max}(\rho)=-\mbox{$3\over d$}\rho w_0''(\rho)$.
The critical temperature is the largest of those,
$T_c=T_{\rm max}(\rho_c)=\mbox{$3\over d$}\rho_cw_0''(\rho_c)$.

The isothermal sound speed $v_T$ readily follows,
\beq
v_T^2\ =\ {\rho\over h}\left({\del p\over\del\rho}\right)_T\
=\ -{\rho\over h}\rho T{|\bold{\sigma}|\over\sigma_{\eps\eps}}\
=\ {\third dT+\rho w_0'' \over\mbox{$5\over6$}dT+w_0'}\ ,\label{vT}
\eeq
where $h\equiv p+\eps$ and 
we have used $(\del p/\del\rho)_T=\del_\rho p_T(\rho)$,
while the isentropic sound speed $v_s$ is given by
\beqar\label{vs}
v_s^2 &=& {\rho\over h}\left({\del p\over\del\rho}\right)_s\
=\ \del_\eps p(\eps,\rho)+{\rho\over h}\del_\rho p(\eps,\rho)\\ \nonumber
&=& -{T\over h}
[h^2\sigma_{\eps\eps}+2h\rho\sigma_{\eps\rho}+\rho^2\sigma_{\rho\rho}]\
=\ {\mbox{$5\over9$}dT+\rho w_0'' \over\mbox{$5\over6$}dT+w_0'}\ ,
\eeqar
where $s=\sigma/\rho$ is the entropy per particle
and we have used that the requirement $\delta s\doteq0$ implies
$\rho\delta\eps\doteq h\delta\rho$.
We note that $v_s^2\geq\ v_T^2$ for $T\geq0$.

\section{Interface}
\label{interface}

Consider a planar interface between two semi-infinite systems and
let the bulk values of the charge and energy densities in system $i$
be $\eps_i$ and $\rho_i$, respectively, and assume that $\rho_1\leq\rho_2$.
Presumably the local  densities $\rho(x)$ and $\eps(x)$ rapidly approach
these asymptotic values away from the interface.
We may generally define the {\em interface location function} \cite{WDM:Book},
\beq
g_\rho(x)\ \equiv\ {\del_x\rho(x)\over\rho_2-\rho_1}\ ,\,\
\int_{-\infty}^{+\infty}\! g_\rho(x)\,dx\ =\ 1\ ,
\eeq
which peaks near the interface and is normalized to unity.
Its moments provide quantitative characteristics of the interface profile
and we list here the first three \cite{WDM:Book},
\beqar
&~&{\rm Location\!:}\,~~~ \bar{x}\ =\ \langle x\rangle\ 
	\equiv\ \int_{-\infty}^{+\infty}\!\!\! x\,g_\rho(x)dx\ ,\\
&~&{\rm Width\!:}~~~~~~~ b\ =\ [\langle(x-\bar{x})^2\rangle]^{1/2}\ ,\\
&~&{\rm Skewness\!:}~~ \gamma_3\ =\ \langle(x-\bar{x})^3\rangle/b^3\ .
\eeqar

In order to extract the interface tension,
we follow the discussion in Ref.\ \cite{MyersNPA436}.
Thus, the difference between the actual diffuse energy density $\eps(x)$ 
and that associated with the corresponding sharp-surface configuration 
that would result if there were no gradient term
is given by 
\beq\label{Es}
 \eps_{12}(x)\
=\ \eps(x)-\eps_i -{\eps_2-\eps_1\over\rho_2-\rho_1}\,(\rho(x)-\rho_i)\ ,\,\
i=1,2\ .
\eeq
The function $\eps_{12}(x)$ is obviously smooth and,
moreover, it tends quickly to zero away from the interface, 
so it is indeed peaked in the interface region.
(It is elementary to see that it does not matter whether one 
uses $i=1$ or $i=2$ in the above expression.)
Its integral is then the total energy deficit associated with the
diffuse interface, the {\em interface tension},
\beq
\gamma_{12}\ \doteq\ \int_{-\infty}^{+\infty}\eps_{12}(x)\, dx\ ,
\eeq
which can thus readily be calculated once the profile densities
$\rho(x)$ and $\eps(x)$ are known.


			\end{document}